\documentclass[runningheads]{llncs}
\usepackage{amsmath,amsfonts}
\usepackage{algorithmic}
\usepackage{array}
\usepackage{textcomp}
\usepackage{stfloats}
\usepackage{url}
\usepackage{verbatim}
\usepackage{graphicx}
\usepackage{balance}
\usepackage{xcolor}
\usepackage[caption=false,font=normalsize,labelfont=sf,textfont=sf]{subfig}
\def\BibTeX{{\rm B\kern-.05em{\sc i\kern-.025em b}\kern-.08em
		T\kern-.1667em\lower.7ex\hbox{E}\kern-.125emX}}
\usepackage{orcidlink}
\usepackage{makecell}
\usepackage{multirow}
\usepackage{booktabs} 
\usepackage{cite}
\usepackage{amsmath,amssymb,amsfonts}
\usepackage{algorithmic}
\usepackage{graphicx}
\usepackage{textcomp}
\usepackage{xcolor}
\usepackage{algorithmic}
\usepackage{algorithm}
\begin{document}
\captionsetup[figure]{labelformat=simple, labelsep=colon, name=Figure}

\title{Neural-network-based Self-triggered Observed Platoon Control for Autonomous Vehicles}

\titlerunning{Self-triggered Observed Platoon Control for AVs}

\author{Zihan Li\inst{1}\orcidlink{0009-0008-9125-1951} \and
Ziming Wang\inst{2,3}\orcidlink{0000-0001-7000-9578} \and
Chenning Liu\inst{1}\orcidlink{0009-0000-9851-0305}  \and Xin Wang$^*$\inst{1}\orcidlink{0000-0003-2070-960X}}

\authorrunning{Z. Li et al.}

\institute{Southwest University, Chongqing, 400715, China \and
Swinburne University of Technology, Melbourne, VIC 3122, Australia
\and
Tsinghua University, Beijing, 100084, China\\
\email{xinwangswu@163.com}}
\maketitle    
\begin{abstract}
This paper investigates autonomous vehicle (AV) platoon control under uncertain dynamics and intermittent communication, which remains a critical challenge in intelligent transportation systems. To address these issues, this paper proposes an adaptive consensus tracking control framework for nonlinear multi-agent systems (MASs). The proposed approach integrates backstepping design, a nonlinear sampled-data observer, radial basis function neural networks, and a self-triggered communication mechanism. The radial basis function neural networks approximate unknown nonlinearities and time-varying disturbances, thereby enhancing system robustness. A distributed observer estimates neighboring states based on limited and intermittent measurements, thereby reducing dependence on continuous communication. Moreover, self-triggered mechanism is developed to determine triggering instants, guaranteeing a strictly positive minimum inter-event time and preventing Zeno behavior. The theoretical analysis proves that all closed-loop signals are uniformly ultimately bounded (UUB), and tracking errors converge to a compact set. Simulation results demonstrate that the proposed approach achieves high robustness, adaptability, and communication efficiency, making it suitable for real-world networked vehicle systems.

\keywords{platoon control  \and self-triggered control \and radial basis function neural networks  \and autonomous vehicles.}
\end{abstract}
\section{Introduction}
Autonomous vehicle platoon control has become a key research direction in intelligent transportation systems, aiming to improve traffic efficiency, reduce fuel consumption, and enhance road safety through coordinated vehicle motion~\cite{lemma1}. Vehicle-to-vehicle (V2V) and vehicle-to-infrastructure (V2I) communications enable real-time information exchange, thereby facilitating the formation and maintenance of stable platoons under predefined communication topologies. However, traditional control strategies often rely on idealized assumptions about vehicle dynamics and communication reliability, overlooking practical issues such as modeling uncertainties~\cite{wang2025fixed,cyber}, unknown disturbances~\cite{case}, and intermittent communication failures~\cite{intro6}. These challenges motivate the development of robust and adaptive frameworks capable of maintaining performance under dynamic conditions.

In practical platoons, packet loss, time-varying delays, and bandwidth constraints are inevitable. To cope with these problems, observer-based designs have been introduced to estimate the unmeasured or unreliable states of neighboring vehicles using limited sensor data. The observer-based self-triggered adaptive control framework reduces dependence on continuous communication, improving robustness and practicality under real-world constraints~\cite{tnnls,ni2024sliding,eve3.2,li2024passivity}. Moreover, integrating adaptive techniques with observer design enables compensation for modeling errors and external disturbances, ensuring accurate estimation and stable cooperative motion.

Traditional time-triggered control schemes execute control tasks periodically at fixed intervals~\cite{eve1,chufa,lemma2}, leading to unnecessarily high-frequency updates and resource consumption. Event-triggered control~\cite{eve4,eve2,cogcomp2025} has been proposed to address this by updating control signals only when specific conditions are violated, significantly reducing communication load while maintaining closed-loop stability. However, ETC typically requires continuous monitoring of the triggering condition, which, despite reducing communication, still imposes a substantial burden on the agent's sensing and computational resources. To overcome this limitation, self-triggered control (STC) has emerged as an advanced strategy. STC eliminates continuous monitoring by calculating the precise next triggering instant $t_{k+1}$ at the current event time $t_k$. This allows the control input to be held constant across the interval $t \in [t_k, t_{k+1})$. Existing research on STC has demonstrated its ability to maintain stability and performance in various multi-agent systems, particularly in tackling complex issues like distributed tracking control and ensuring resource efficiency under strict network constraints~\cite{eve5,lemma2,tnnls,li2024passivity}. This prediction-based update mechanism drastically enhances the practicality of implementing cooperative control in resource-constrained environments.

Motivated by these considerations, it is with this goal that the work, specifically this paper, proposes an adaptive consensus tracking framework for nonlinear high-order multi-agent systems (MASs) based on backstepping, filtering techniques, radial basis function neural networks, and distributed observers. The main contributions are summarized as follows:
\begin{figure}[htbp]
    \centering
    \includegraphics[width=0.75\linewidth]{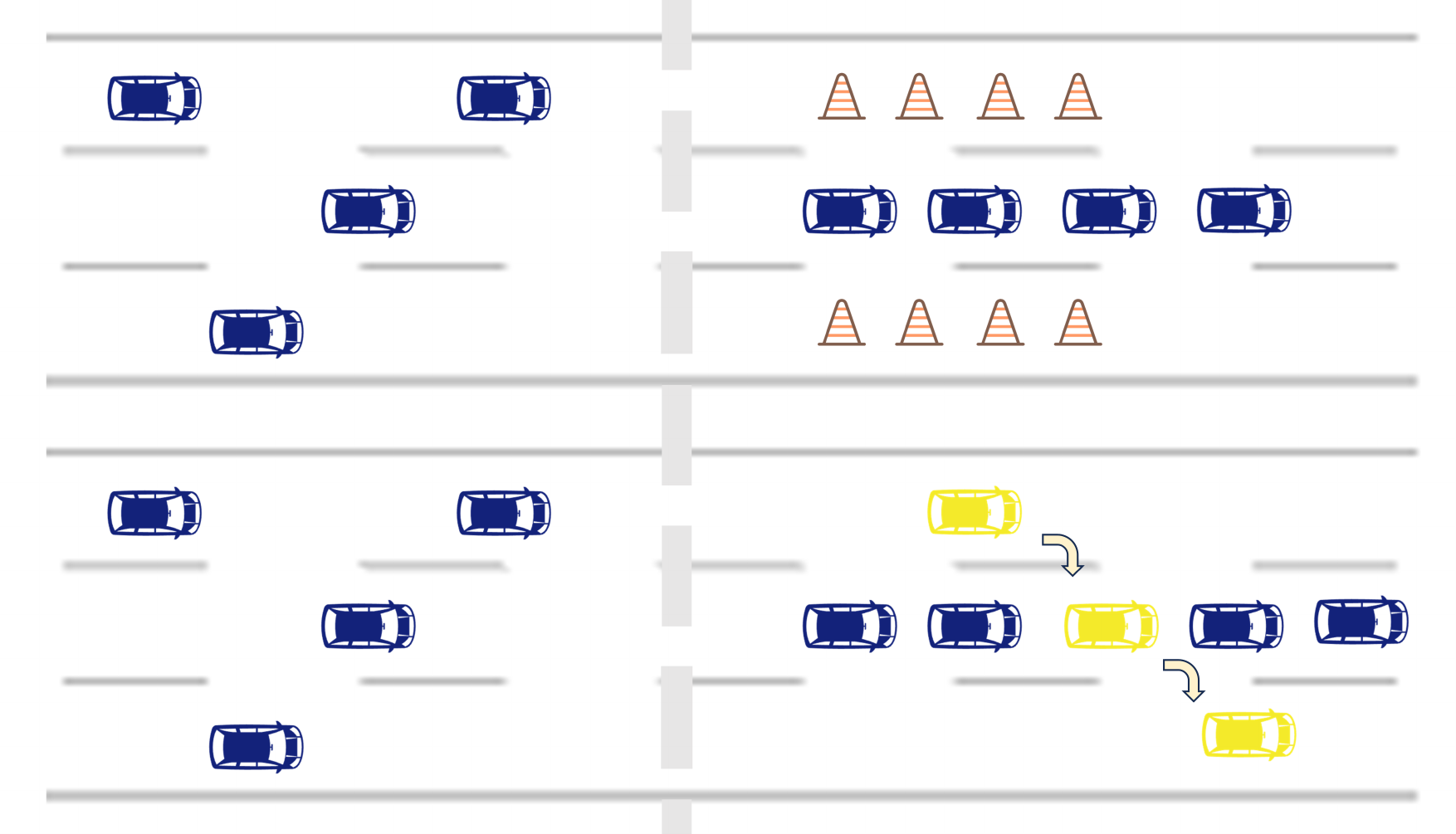}

    \caption{AVs formation control in linear formation and linear-queue formation.}
    \label{fig:placeholder}
\end{figure}

\begin{itemize}
    \item Adaptive neural networks compensation: neural networks are employed to approximate and compensate for unknown nonlinearities and disturbances, enhancing robustness and adaptability without requiring precise model knowledge.
    
    \item A distributed observer estimates neighboring agents’ full states using limited and intermittent output data, reducing reliance on continuous communication and perfect measurements~\cite{tnnls,eve3.1}.
    
    \item A self-triggered control mechanism determines the next triggering instant based solely on the current control signal and its rate of change, avoiding continuous state monitoring~\cite{eve5,lemma2}. It guaranties a strictly positive minimum inter-event time and allows adjustable trade-offs between control accuracy and communication burden, enhancing practicality for networked systems.
\end{itemize}

The remainder of this paper is organized as follows. Section~\ref{sec:preliminary} reviews preliminaries on MASs, observer design, and graph theory, and formulates the problem. Section~\ref{sec:method} presents the self-triggered control mechanism. Section~\ref{sec:stability} analyzes stability. Section~\ref{sec:example} provides simulation results. Finally, Section~\ref{sec:conclusion} concludes the paper.

\section{Preliminaries and Problem Formulation}\label{sec:preliminary}

\subsection{Graph Theory}

In the AV platoon, communication interactions are modeled by a directed graph $\mathcal{G} = (\mathcal{V}, \mathcal{E})$, where $\mathcal{V}$ is the set of follower vehicles and $\mathcal{E}$ represents the directed communication links. The communication topology is described by the adjacency matrix $\mathcal{A} = [a_{ij}] \in \mathbb{R}^{N \times N}$, where $a_{ij} = 1$ if follower $i$ receives information from vehicle $j$, and $a_{ij} = 0$ otherwise. The in-degree matrix $\mathcal{D} = \mathrm{diag}\{d_i\}$ satisfies $d_i = \sum_{j=1}^{N} a_{ij}$, and the Laplacian matrix is $\mathcal{L} = \mathcal{D} - \mathcal{A}$.
The leader-follower coupling is defined by the pinning gain matrix $\mathcal{B} = \mathrm{diag}\{b_i\}$, where $b_i = 1$ if the follower $i$ can access the leader’s state and $b_i = 0$ otherwise. The matrix $\mathcal{H} = \mathcal{L} + \mathcal{B}$ plays a key role in stability analysis; if the communication graph contains a directed spanning tree rooted at the leader, all eigenvalues of $\mathcal{H}$ possess positive real parts. Notably, the proposed control scheme does not require explicit knowledge of $\mathcal{H}$, instead, it relies only on locally sampled information, thereby enhancing robustness against communication uncertainties.

\subsection{Systems Model}
	The longitudinal dynamics of each autonomous vehicle (AV) in the platoon are considered. The tenth vehicle is described by a second-order nonlinear system:
  \begin{equation}
\dot{p}_{i} = v_{i}
\end{equation}
\begin{equation}
\dot{v}_{i} = \frac{1}{m_{i}} \left( F_{i} - f_{i}(v_{i}) \right) + d_{i}\label{eq:vehicle_dynamics_velocity}
\end{equation}
where $i\in \mathcal{N}={1,2,\dots,N}, p_{i} \in \mathbb{R} $ and $ v_{i} \in \mathbb{R} $ represent the position and velocity of the vehicle, respectively $ m_{i} > 0 $ denotes vehicle mass, and $ F_{i} \in \mathbb{R} $ is the control input force. The term $ f_{i}(v_{i}) $ encapsulates unknown non-linear resistance forces, including aerodynamic drag and rolling resistance. The term $ d_{i} \in \mathbb{R} $ represents unknown external disturbances.

	\subsection{Radial Basis Function Neural Networks}
    
According to research~\cite{rela2,tsmc2025}, radial basis function neural networks are employed to approximate the unknown nonlinearities \(\Delta_i(v_i, d_i)\) inherent in vehicle dynamics, leveraging their universal approximation capability to model any continuous function \(F({Z}) ={W}^{*T} {\Psi}({Z}) + \epsilon({Z})\) over a compact set. Here, \({Z}\) denotes the input vector, \({W}^*\) represents the ideal weight vector, \({\Psi}({Z})\) consists of a set of Gaussian basis functions \(\psi_k({Z}) = \exp\left( -{\|{Z} -{\mu}_k\|^2}/{\eta_k^2} \right)\), and \(\epsilon({Z})\) signifies the bounded approximation error. The online estimate of the uncertainty is constructed as \(\hat{\Delta}_i({Z}_i) = \hat{{W}}_i^T {\Psi}_i({Z}_i)\), where \(\hat{{W}}_i\) is the adaptive estimate of the ideal weights, and \(\tilde{{W}}_i = {W}_i^* - \hat{{W}}_i\) denotes the weight estimation error. This approach facilitates the development of robust and self-learning control capabilities without reliance on precise prior models.

\subsection{Self-triggered Mechanism Design}
In self-triggered control, the control input is held constant between consecutive triggering instants and is updated at the current trigger time $t_k$. The control hold law is given by:
\begin{equation}
u_i(t) = u_i(t_k), \quad \forall t \in [t_k, t_{k+1})\label{eq:hold_law}
\end{equation}

The core of this mechanism involves computing the next triggering instant $t_{k+1}$ at the current time $t_k$. The proposed triggering control mechanism is designed as follows:

\begin{equation}
t_{k+1} = t_k + \min\left\{ \frac{ s_\sigma |u_i(t_k)| + s_D }{ \max( |\dot{\hat{u}}_i(t_k)|, s_\Lambda ) },\ T_{\mathrm{max}} \right\}\label{eq:self_trigger_rule}
\end{equation}
where $T_{\mathrm{max}}$ specifies the maximum inter-event time to ensure the minimum update frequency. Constants include: $s_\sigma \in (0,1)$ scaling the threshold with control magnitude; $s_D > 0$ providing a fixed threshold to prevent unbounded intervals; $s_\Lambda > 0$ maintaining a minimum update rate when $\dot{\hat{u}}_i(t_k)$ is small. In contrast to event-triggered control, this self-triggered control mechanism (\ref{eq:self_trigger_rule}) eliminates the need for continuous state measurement. The computation is executed only at each triggering instant, thereby reducing the agent’s sensing workload.

\section{Main Result}\label{sec:method}
    
To facilitate the controller design, the model is rewritten by defining the control input as $ u_{i} = F_{i}/m_{i} $. The uncertainty, which includes internal resistance and external disturbances, is denoted as $ \Delta_{i} = -f_{i}(v_{i})/m_{i} + d_{i} $. With these new definitions, the vehicle dynamic model can therefore be transformed into the following presented compact form:

\begin{equation}
\begin{array}{c}
\dot{p}_{i}=v_{i} \\
\dot{v}_{i}=u_{i}+\Delta_{i}\left(v_{i}, d_{i}\right)\label{eq:compact_dynamics_model}
\end{array}
\end{equation}

In this paper, it is the uncertainty $ \Delta_{i}(\cdot) $ that is an unknown non-linear function. To handle this, neural network is employed to approximate it over a compact set $ \Omega $ with arbitrary accuracy, leading to the following expression:
\begin{equation}
\Delta_{i}=W_{i}^{* T} \Psi_{i}\left(Z_{i}\right)+\epsilon_{i}\left(Z_{i}\right)\label{eq:rbf_nn_approximation}
\end{equation}
where $Z_{i} = [v_{i}, d_{i}]^{T} \in \Omega \subset \mathbb{R}^{2}$ is the input vector to the neural networks, $W_{i}^{*} \in \mathbb{R}^{\ell}$ is the ideal optimal weight vector, and $\Psi_{i}(Z_{i}) = [\psi_{i 1}(Z_{i}), \psi_{i 2}(Z_{i}), \ldots, \psi_{i\ell}(Z_{i})]^{T} \in \mathbb{R}^{\ell}$ is the Gaussian basis function vector. The term $\epsilon_{i}(Z_{i})$ denotes the bounded approximation error satisfying $|\epsilon_{i}(Z_{i})| \leq \bar{\epsilon}_{i}$, where $\bar{\epsilon}_{i} > 0$ is a designed constant. Substituting (\ref{eq:rbf_nn_approximation}) into (\ref{eq:compact_dynamics_model}) yields the final model used for controller synthesis:
\begin{equation}
\begin{array}{l}
\dot{p}_{i}=v_{i} \\
\dot{v}_{i}=u_{i}+W_{i}^{* T} \Psi_{i}\left(Z_{i}\right)+\epsilon_{i}\left(Z_{i}\right)\label{eq:augmented_dynamics_with_nn}
\end{array}
\end{equation}

It is crucial to note that the control algorithm developed in this paper does not require precise knowledge of the vehicle mass $ m_{i} $, the nonlinear function $ f_{i}(v_{i}) $, or the disturbance $ d_{i} $. Instead, the adaptive laws and the neural network estimator will identify and compensate for the uncertainty $ \Delta_{i} $, enhancing the system's robustness and adaptability.

Given the limitations and the need to reduce communication load, a sampling-based state observer is proposed. It estimates the full state of both the preceding and ego vehicles using sampled position data. Let $t_k$ denote the $k$-th sampling instant with fixed interval $T = t_{k+1} - t_k$. The measured position of the preceding vehicle $(i-1)$, affected by sensor noise and perception errors, is given as:

\begin{equation}
\tilde{p}_{i-1}(t_k) = p_{i-1}(t_k) + \zeta_{i-1}(t_k)\label{eq:preceding_vehicle_measurement}
\end{equation}
where \( \zeta_{i-1}(t_k) \) is the bounded measurement noise that satisfies \( | \zeta_{i-1}(t_k) | \leq \bar{\zeta}_{i-1} \).A Luenberger-type nonlinear sampled-data observer is proposed for the ego vehicle \( i \) to estimate the unmeasurable states:
\begin{align}
\dot{\hat{p}}_{i}&=\hat{v}_{i}+L_{i1}(\tilde{p}_{i-1}(t_k)-\hat{p}_{i}) \label{eq:luenberger_observer_p} \\
\dot{\hat{v}}_{i}&=u_{i}+\hat{\Delta}_{i}(\hat{v}_{i})+L_{i2}(\tilde{p}_{i-1}(t_k)-\hat{p}_{i}) \label{eq:luenberger_observer_v}
\end{align}

Here, \( \hat{p}_i \) and \( \hat{v}_i \) are the estimated positions and velocities of the ego vehicle, respectively. \( L_{i1} > 0 \) and \( L_{i2} > 0 \) are the observer gain constants to be designed. \( \hat{\Delta}_i(\hat{v}_i) = \hat{W}_i^T \Psi_i(\hat{v}_i) \) is the online estimate of the lumped uncertainty of the neural network.The dynamics of the observer error are defined as \( e_{p_i} = p_i - \hat{p}_i \) and \( e_{v_i} = v_i - \hat{v}_i \). 

\textbf{Assumption 1.} Based on research\cite{arxiv,trc}, the vehicle velocity dynamics are Lipschitz continuous, meaning there exists a designed constant \( \gamma_i > 0 \) such that:
\begin{equation}
| \dot{v}_i(t_1) - \dot{v}_i(t_2) | \leq \gamma_i | t_1 - t_2 |, \quad \forall t_1, t_2 \in [t_k, t_{k+1})\label{ass:1}
\end{equation}

\textbf{Assumption 2.} Based on research\cite{arxiv,trc}, the rate of change of vehicle position is bounded; that is, the velocity satisfies \( | v_i(t) | \leq v_{i}^{\max} \) for all \( t \geq 0 \), where \( v_{i}^{\max} > 0 \) is an unknown constant.

Under these assumptions and with appropriately chosen gains \( L_{i1}, L_{i2} \), it is guaranteed that the dynamics of the observation error are ultimately uniformly bounded. The key innovation lies in the structure of the observer, which uses the sampled position of the preceding vehicle $\tilde{p}_{i-1}(t_k) $as a corrective input rather than requiring continuous state information from other vehicles or infrastructure.

In control design, the formulation of the tracking error is central to achieving control objectives. For bipartite tracking control with competitive-cooperative relationships, the definition of the tracking error differs from that of traditional consensus. The bipartite tracking error $ e_{i1} $  for agent $ i $  is typically defined as:
\begin{equation}
e_{i1}=\sum_{j=1}^{N}|a_{ij}|(y_{i}-\operatorname{sgn}(a_{ij})y_{j})+|b_{i}|(y_{i}-\operatorname{sgn}(b_{i})y_{r})\label{ass:2}
\end{equation}
where $ a_{ij} $ is the element of the adjacency matrix representing the cooperative-competitive relationship between agents, $ b_{i} $ represents the connection to the leader, $ y_{i} $ and $ y_{j} $ are the outputs of the agents, $ y_{r} $ is the output of the leader, and $ \operatorname{sgn}(\cdot) $ is the signum function. This formulation ensures that, under competitive interactions, the agents asymptotically converge to values with identical magnitudes and opposite signs.

In vehicle formation control, the tracking error is more directly related to the system states and the desired trajectory. For agent $i$, its position tracking error $ z_{i,1} $ and velocity tracking error $ z_{i,2} $ are typically defined as: 

\begin{align}
z_{i,1} &= \hat{x}_{i}-q_{i} \\
z_{i,2} &= \hat{v}_{i}-\dot{q}_{i}-\alpha_{i}\label{ass:3}
\end{align}

Here,  $ \hat{x}_i $ and  $ \hat{v}_i $ represent the position and velocity of agent i estimated by the observer, $ q_i $ and $ \dot{q}_i $ denote its desired position and velocity trajectories, and $ \alpha_i $ is the virtual control law designed in the backstepping procedure.

\section{Stability Analysis}\label{sec:stability}
\itshape Theorem 1:
\upshape Consider the MASs (\ref{eq:compact_dynamics_model}) with uncertainties. With the self-triggered mechanism defined by  (\ref{eq:hold_law}) and (\ref{eq:self_trigger_rule}), the following properties hold:
\begin{enumerate}
    \item All signals in the closed-loop system, including the tracking error ${e}$, the neural network weight estimates $\hat{{W}}$, and the disturbance constant estimates $\hat{{\Theta}}$, are uniformly ultimately bounded (UUB).
    \item The tracking error ${e}$ converges to a compact set around the origin, the size of which can be adjusted by the design constants $s_\sigma$, $s_D$, $s_\Lambda$, and the adaptive gains.
    \item The inter-event times are strictly positive, i.e., there exists a constant $\tau > 0$ such that $\inf_k (t_{k+1} - t_k) \geq \tau$; thus, Zeno behavior is excluded.
\end{enumerate}

\itshape Proof:
\upshape Consider the Lyapunov function candidate:
\begin{equation}
\label{eq:lyapunov}
V = \frac{1}{2} {e}^T {P} {e} + \frac{1}{2} \left(\tilde{{W}}^T {\Gamma}^{-1} \tilde{{W}}\right) + \frac{1}{2} \tilde{{\Theta}}^T {\Xi}^{-1} \tilde{{\Theta}}
\end{equation}
where ${e}$ is the tracking error vector, $\tilde{{W}}$ and $\tilde{{\Theta}}$ are the weight estimation error matrices for the neural network and the disturbance constants, respectively, and ${P}$, ${\Gamma}$, ${\Xi}$ are positive definite matrices. The time derivative of $V$ along the trajectories of the system is given by:
\begin{equation}
\dot{V} = {e}^T {P} \dot{{e}} + \left(\tilde{{W}}^T {\Gamma}^{-1} \dot{\tilde{{W}}}\right) + \tilde{{\Theta}}^T {\Xi}^{-1} \dot{\tilde{{\Theta}}}
\end{equation}

Substituting the system dynamics, the control law $u_i(t) = u_i(t_k)$, and the adaptive update laws, we obtain an expression that contains a negative definite term $- {e}^T {Q} {e}$ and cross terms involving the control input error $e_u(t) = u_i(t_k) - \bar{u}_i(t)$, where $\bar{u}_i(t)$ denotes the ideal continuous control signal. The design of the self-triggered mechanism ensures that, for $t \in [t_k, t_{k+1})$, the control input error $e_u(t)$ satisfies
\begin{equation}
\label{eq:error_bound}
|e_u(t)| \leq s_\sigma |u_i(t_k)| + s_D
\end{equation}
where $s_\sigma$ and $s_D$ are positive design constants.            
Using Young's inequality and standard norm inequalities, the derivative $\dot{V}$ can then be upper bounded as:
\begin{align}
\dot{V} &\leq - \lambda_{\min}({Q}) \|{e}\|^2 + c_1 \|{e}\| \cdot |e_u(t)| + c_2 \nonumber \\
        &\leq - \alpha V + \beta
\end{align}

A key step in the analysis is to bound the control input error.  
The self-triggered mechanism ensures that, for $t \in [t_k, t_{k+1})$, the control input error $e_u(t)$ satisfies
\begin{equation}
\label{eq:error_bound}
|e_u(t)| \leq s_\sigma |u_i(t_k)| + s_D
\end{equation}
where $s_\sigma$ and $s_D$ are positive design constants. Using Young’s standard norm properties, the derivative of the Lyapunov function $V$ can be upper bounded as
\begin{align}
\dot{V} &\leq - \lambda_{\min}(Q) \|e\|^2 + c_1 \|e\| \cdot |e_u(t)| + c_2 \nonumber \\
        &\leq - \alpha V + \beta
\end{align}
for some positive constants $c_1, c_2, \alpha, \beta > 0$. $\alpha$ characterizes the exponential convergence rate of the closed-loop system, indicating how rapidly the tracking and observer errors decay in the absence of triggering-induced accumulation. In contrast, $\beta$ quantifies the residual steady-state bound resulting from sampling effects, estimation inaccuracies, and triggering-induced input perturbations. Then integrating both sides gives
\begin{equation}
\label{eq:Vbound}
0 \leq V(t) \leq \frac{\beta}{\alpha} + \left(V(0) - \frac{\beta}{\alpha}\right)e^{-\alpha t},
\end{equation}
which shows that $V(t)$ exponentially converges to a bounded residual set of radius.

Therefore, according to the properties of positive definite matrices, it is corroborated that the observer error $e$, tracking errors $z_{i,1}$ and $z_{i,2}$, and constant estimation errors $\tilde{W}_{i,j}$ and $\tilde{\sigma}_i$ are all bounded, i.e.,
\[
\|e\|^2 \leq \frac{2V}{\phi_{\min}(P)}, \quad
\|z_{i,1}\|^2 \leq 2V, \quad
\|z_{i,2}\|^2 \leq 2V
\]
\[
\|\tilde{W}_{i,j}\|^2 \leq \frac{2V}{\phi_{\min}(\mathcal{O}_{i,j})}, \quad
\|\tilde{\sigma}_i\|^2 \leq \frac{2V}{\phi_{\min}(\Delta_{i,j})}
\]

Now we show that there exists a positive constant $t^* > 0$ such that for all $k \in \mathbb{Z}^+$, the triggering intervals satisfy $\{t_{k+1} - t_k\} \geq t^*$.  
To this end, by recalling $e_i(t) = w_i(t) - u_i(t)$, during the period $t_i^k \le t_i < t_i^{k+1}$, we obtain
\begin{equation}
\frac{d}{dt}\|e_i\| = \frac{d}{dt}\!\left(e_i^T e_i\right)^{\frac{1}{2}}
= [\operatorname{sgn}(e_{i,1}), \operatorname{sgn}(e_{i,2})]^T e_i = \|w_i\|
\label{eq:ei_derivative}
\end{equation}
where $e_{i,j}$ is the $j$-th component of the vector $e_i$, $j=1,2$.  

This result indicates that the growth rate of the control error is bounded by the control magnitude, thereby ensuring a strictly positive minimum inter-event interval $t^*$ and excluding Zeno behavior.  Hence, all closed-loop signals remain bounded, and overall system stability is guaranteed.  Finally, the inequality $\dot{V} \leq -\alpha V + \beta$ proves that all signals in the closed-loop system, including the tracking error ${e}$ and the estimation errors $\tilde{{W}}$, $\tilde{{\Theta}}$, are UUB.

To exclude Zeno behavior, we prove the existence of a positive lower bound for the inter-event times. The time derivative of the control signal $\bar{u}_i(t)$ can be shown to be bounded. Suppose there exists a constant $\Upsilon > 0$ such that:
\begin{equation}
\left| \frac{d}{dt} \left(\bar{u}_i(t)\right) \right| \leq \Upsilon
\end{equation}

The error $e_u(t)$ evolves from $0$ at $t_k$ to the threshold $s_\sigma |u_i(t_k)| + s_D$ at $t_{k+1}$. The time required for the process of this growth is at least:
\begin{equation}
\tau = \frac{ s_\sigma |u_i(t_k)| + s_D }{ \Upsilon } > 0
\end{equation}

Therefore, the inter-event time is bounded below by a positive constant, $t_{k+1} - t_k \geq \tau > 0$. This result explicitly excludes Zeno behavior,as the bounded rate of change of the ideal continuous control law $\bar{u}_i(t)$ ensures that the time required for the control error to grow from zero to the triggering threshold possesses a positive lower bound $\tau$, thereby guaranteeing a strictly positive interval between consecutive triggering instants.

\section{ILLUSTRATIVE EXAMPLE}\label{sec:example}
\subsection{Simulation Setup and Scenario Design}

To validate the proposed observer-based self-triggered adaptive control mechanism, numerical simulations are conducted in MATLAB for a heterogeneous vehicle platoon comprising one leader and three followers with distinct masses $m_i = [1800, 1900, 1850, 1760]\,\mathrm{kg}$. The simulation uses a sampling interval of $T = 0.001\,\mathrm{s}$ over a $50\,\mathrm{s}$ duration.

Each vehicle $i \in \{1,2,3,4\}$ (vehicle 1 as leader) has a longitudinal-lateral position $p_i = [p_{xi}, p_{yi}]^\top$ and velocity $v_i = [v_{xi}, v_{yi}]^\top$, with control inputs $u_{xi}, u_{yi}$ resisting unknown nonlinear forces $f_{xi}, f_{yi}$ and small disturbances $d_{xi}, d_{yi}$. The dynamics follow:
\[
 \dot{v}_{xi} = (u_{xi} + f_{xi} + d_{xi})/m_i, 
\quad \dot{v}_{yi} = (u_{yi} + f_{yi} + d_{yi})/m_i.
\]

Defining $x_i = [p_{xi}, p_{yi}]^\top$ and $\dot{x}_i = v_i$, the observer estimates $\hat{x}_i = [\hat{p}_{xi}, \hat{p}_{yi}]^\top$ and $\hat{v}_i = [\hat{v}_{xi}, \hat{v}_{yi}]^\top$.

To introduce nonzero initial estimation errors, heterogeneous initial states are assigned. For illustration, the leader is initialized at $x_1(0) = [50,\,6.0]^\top\!\mathrm{m}$ with $v_1(0) = [15,\,0]^\top\!\mathrm{m/s}$, while its observer estimates start from $\hat{x}_1(0) = [48,\,5.5]^\top\!\mathrm{m}$ and $\hat{v}_1(0) = [14,\,0.2]^\top\!\mathrm{m/s}$. The remaining followers are positioned sequentially behind the leader, with comparable longitudinal velocities and similarly perturbed observer initializations. This ensures a non-negligible initial mismatch between the true and estimated states for evaluating observer convergence and self-triggered control performance.

The leader's longitudinal velocity $v_{x1}(t)$ follows: $12\,\mathrm{m/s}$ for $0 \le t < 25$; a smooth S-curve deceleration to $6\,\mathrm{m/s}$ during $25 \le t < 31$; and $6\,\mathrm{m/s}$ for $t \ge 31$. The lateral position remains constant at $p_{y1}(t) = 5.4\,\mathrm{m}$. Controller constants are chosen as \(K_1 = \mathrm{diag}(1.5,1.0)\), \(K_2 = \mathrm{diag}(35,5)\), \(C_1 = C_2 = 8\). 
The neural network employs \(l = 5\) basis functions with centers in \([-12,12]\) and width \(\sigma = 2.5\). 
The self-triggered mechanism involves several constants \((s_\sigma, s_D, s_\Lambda, T_{\max})\), which balance control accuracy and communication efficiency. 
Specifically, \(s_\sigma\) adjusts the triggering sensitivity relative to the control magnitude (smaller values yield higher accuracy but require more updates), while \(s_D\) prevents excessive triggering when the control signal is small, typically set as \(s_D \approx 0.01 u_{\max}\). 
The term \(s_\Lambda\) ensures numerical stability when the control rate is nearly constant and can be chosen based on noise levels. 
The upper bound \(T_{\max}\) avoids overly long inter-event times and is selected according to system dynamics, satisfying \(T_{\max} < 1/\gamma_i\). 
These constants are obtained through analytical constraints and limited simulation tuning, ensuring both practical implementability and theoretical consistency.

Two simulation scenarios are designed to evaluate the performance of the proposed self-triggered control algorithm.

In linear formation scenario, vehicles maintain the same lane with a desired inter-vehicle spacing of \(10\,\mathrm{m}\) behind their immediate predecessor. This setup represents a standard highway cruising condition with longitudinal coordination.

In linear-queue scenario, a gap twice the standard inter-vehicle distance is reserved between the second and third vehicles to allow surrounding traffic to change lanes safely. This prevents unnecessary road occupation and reduces interference with vehicles merging from either side, while testing the controller’s ability to maintain formation under such a specific spacing constraints. Communication efficiency is quantified using the triggering ratio:
\[
\mathrm{Triggering\ Ratio} = 
\frac{N_{\mathrm{trigger}}}{N_{\mathrm{total}}} \times 100\%
\]
where \(N_{\mathrm{trigger}}\) is the number of control updates for each AV, and \(N_{\mathrm{total}}\) is the total simulation step count.  
\begin{table}[H]
    \centering

    \begin{tabular}{cccc}
    \hline
    Vehicle & Self-Triggered & Continuous & Ratio (\%)\\
    \hline
    AV1 & 18185& 50000 & 63.63\\
    AV2 & 44405& 50000 & 11.19\\
    AV3 & 44439& 50000 & 11.12\\
    AV4 & 44702& 50000 & 10.60\\
    \hline
    \end{tabular}
        \caption{Self-triggered control statistics}
    \label{tab:triggering_ratio}
\end{table}
The communication update counts for each vehicle under both the self-triggered and continuous control modes, along with the calculated triggering ratios, are summarized in Table~\ref{tab:triggering_ratio}. To improve visibility, only a subset of the triggering instants is displayed in Figure  \ref{Interval}; the statistical results in Table~\ref{tab:triggering_ratio} are based on the full triggering data. The results demonstrate that the self-triggered mechanism significantly reduces the number of required communication updates compared to continuous control.

\subsection{Simulation Results and Analysis}

\begin{figure*}
            \centering
            \subfloat{\includegraphics[width=0.333\linewidth]{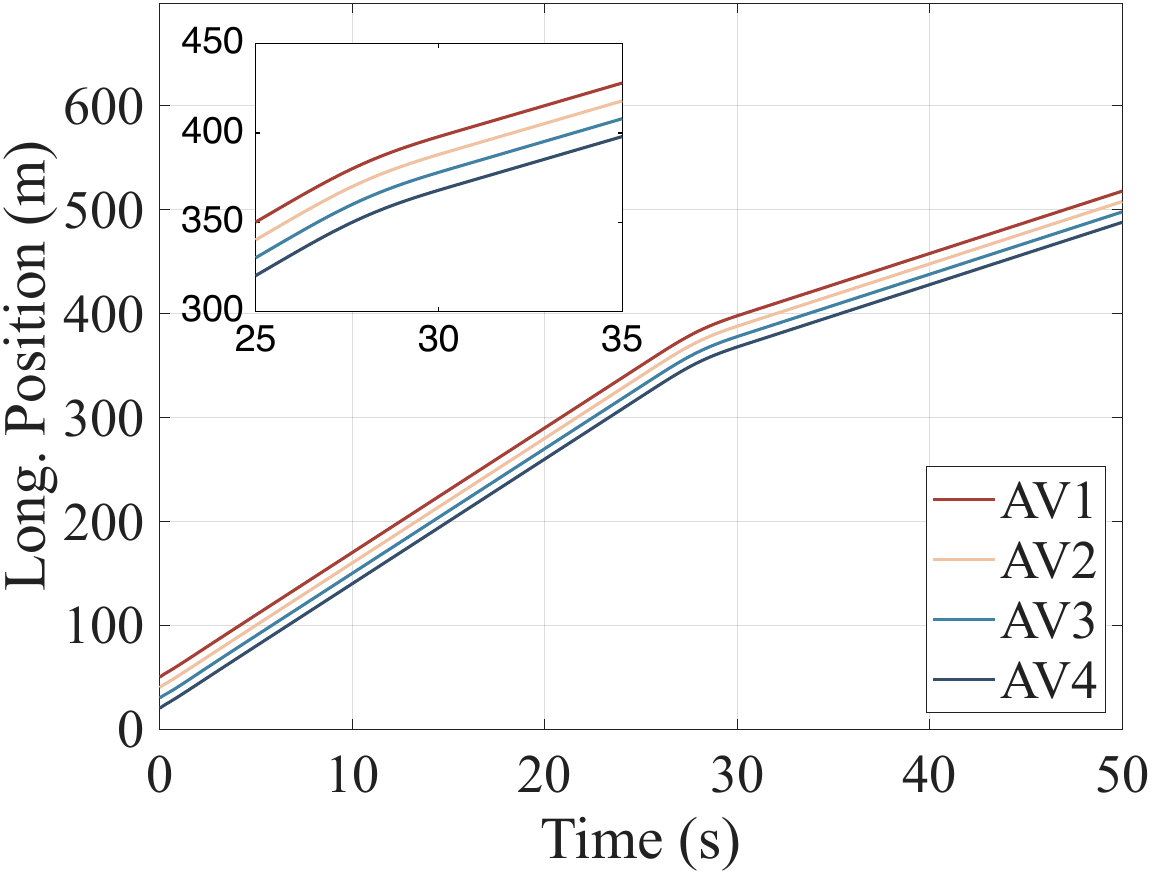}}
            \subfloat{\includegraphics[width=0.333\linewidth]{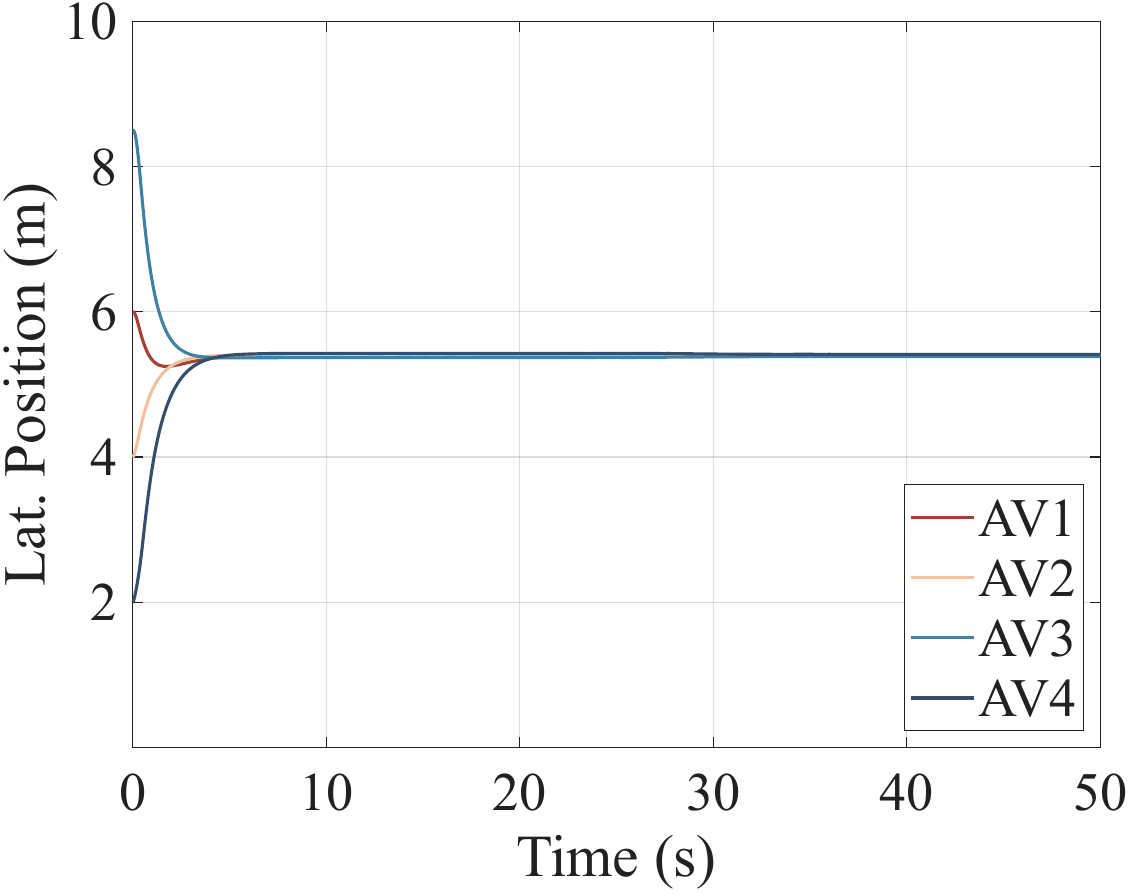}}
            \subfloat{\includegraphics[width=0.333\linewidth]{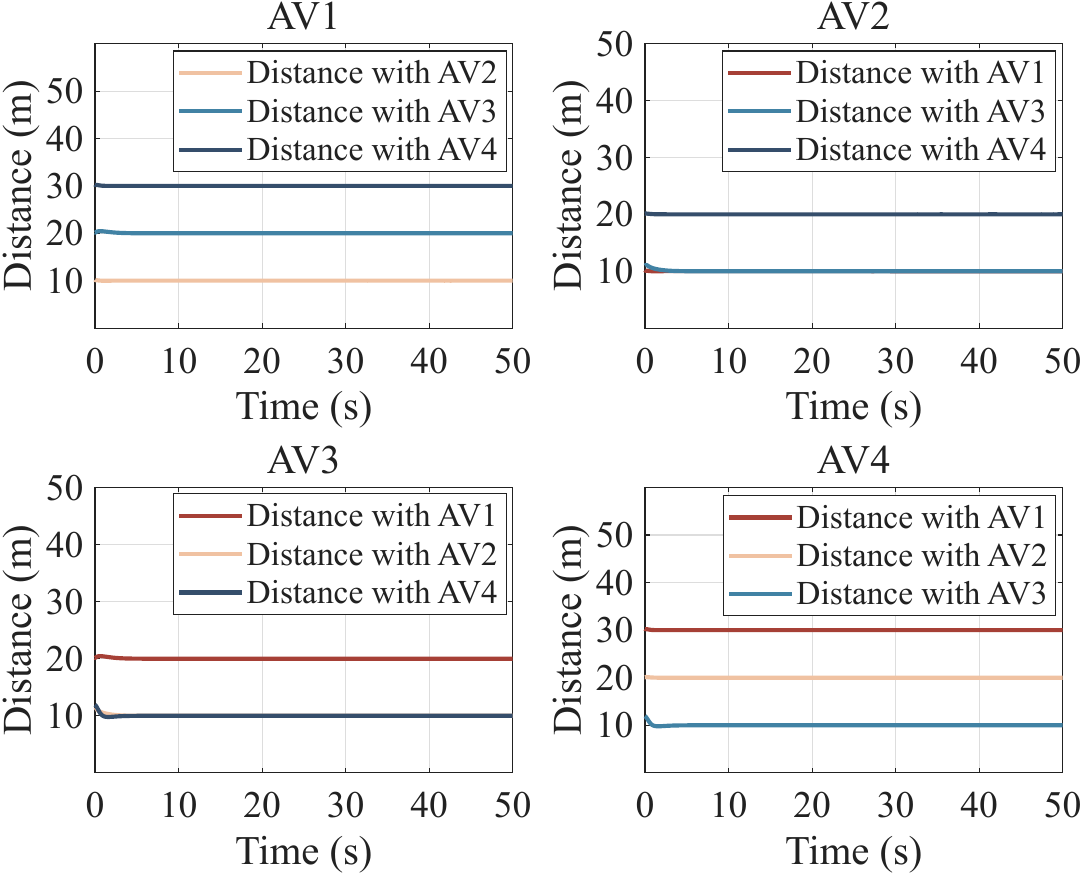}}\\
            \setcounter{subfigure}{0}
            \subfloat[Longitudinal tracking]{\includegraphics[width=0.333\linewidth]{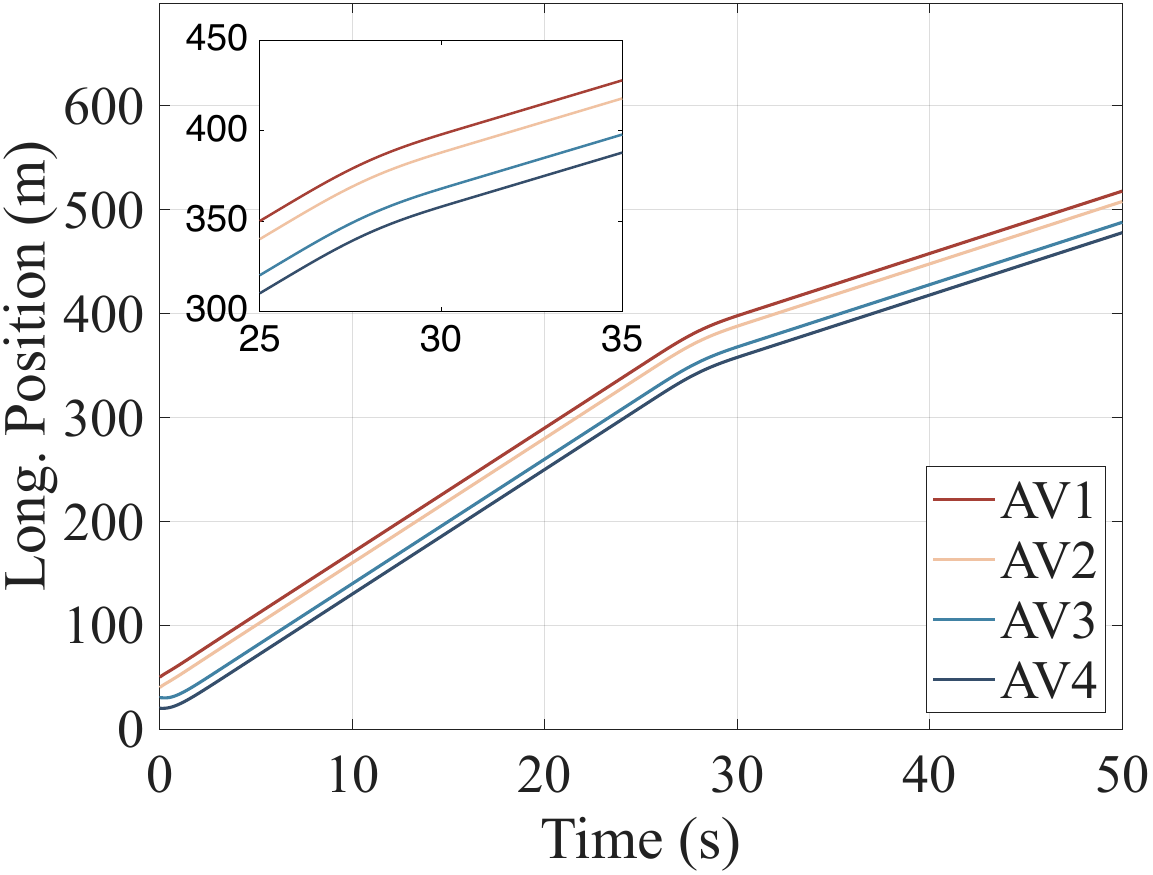}}
            \subfloat[Lateral tracking]{\includegraphics[width=0.333\linewidth]{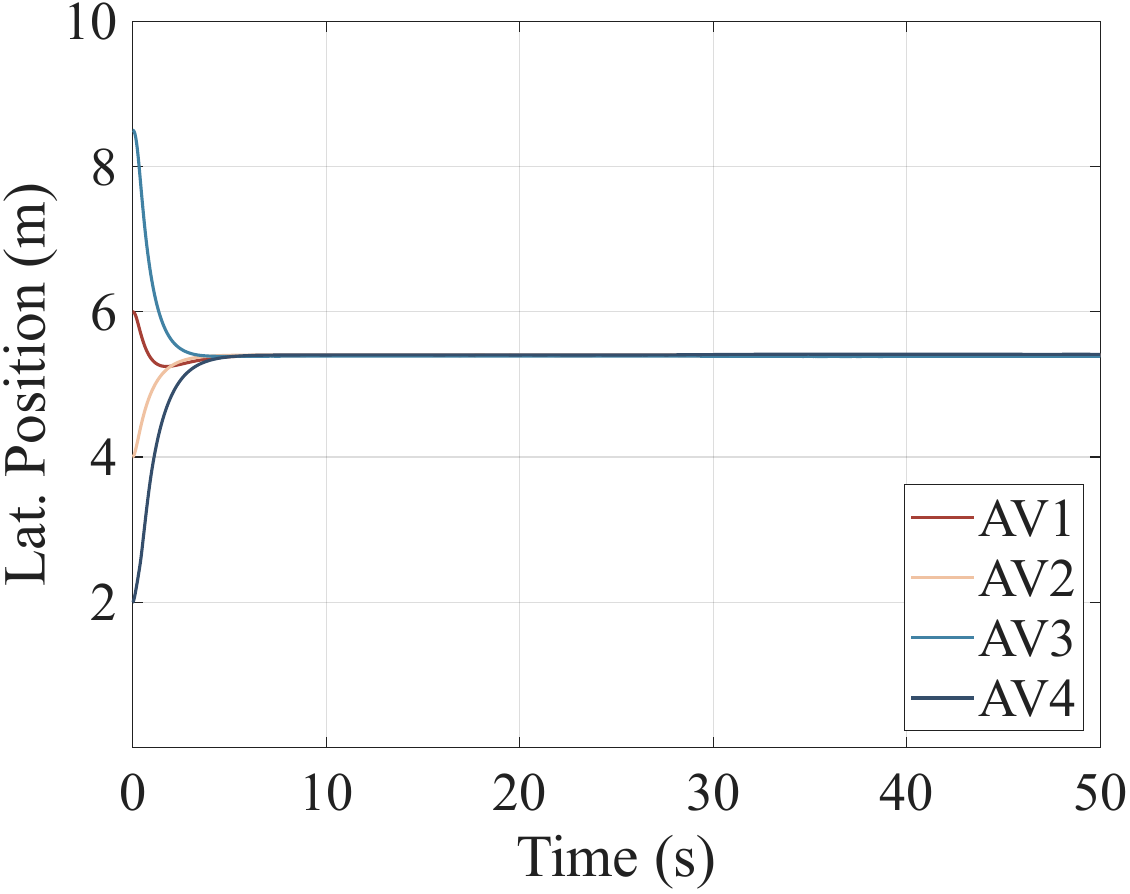}}
            \subfloat[Safe distances]{\includegraphics[width=0.333\linewidth]{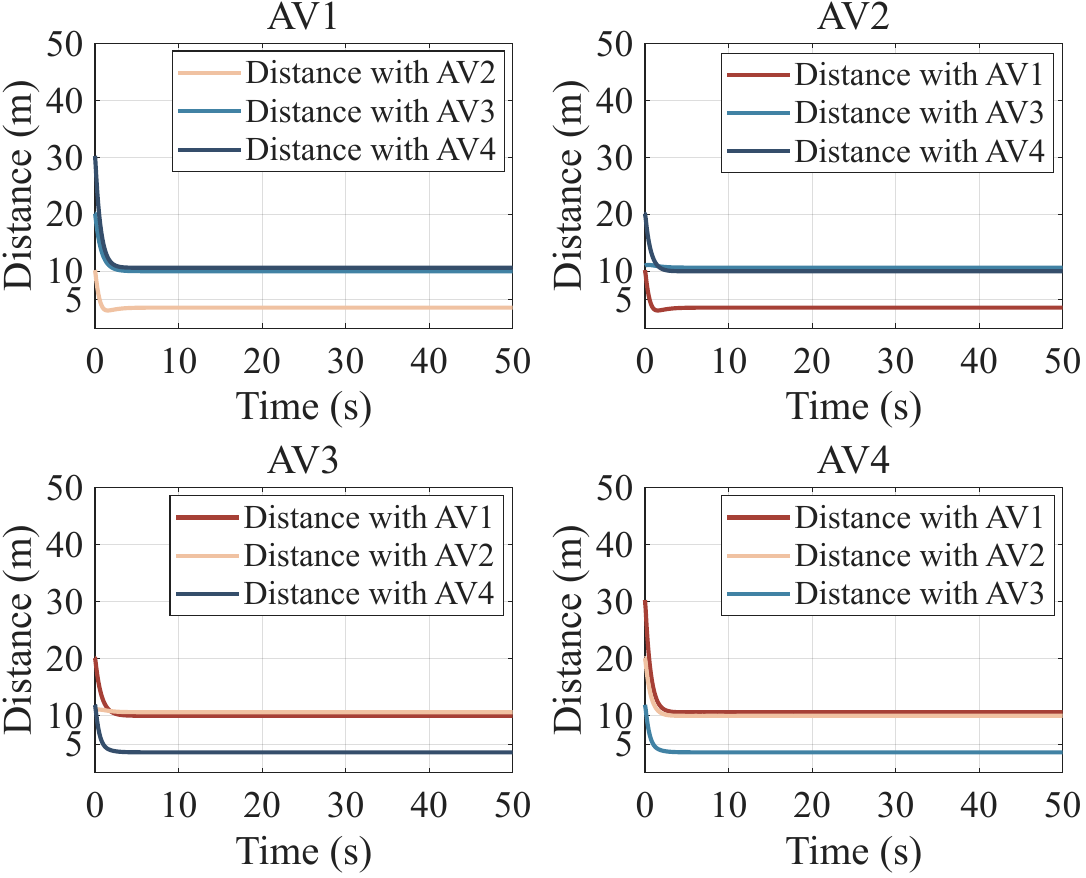}}
            \caption{Longitudinal and lateral tracking performance of the vehicle formation control and safe distances in control process.}
            \label{tracking}
        \end{figure*}


From the longitudinal position plots in Figure~\ref{tracking} (a), the followers respond with minimal delay, preserving the target 10~m spacing with minor fluctuations. The smooth transitions and bounded spacing errors confirm that the PD control with adaptive neural network compensation suppresses overshoot and enhances disturbance rejection. The lateral trajectories in Figure~\ref{tracking} (b), the linear formation maintains nearly constant lateral offsets, while in the linear-queue formation, followers converge smoothly to the leader's path within 15~s without overshoot, indicating well-damped and stable lateral dynamics.

\begin{figure}[htbp]
    \centering
    \includegraphics[width=0.75\linewidth]{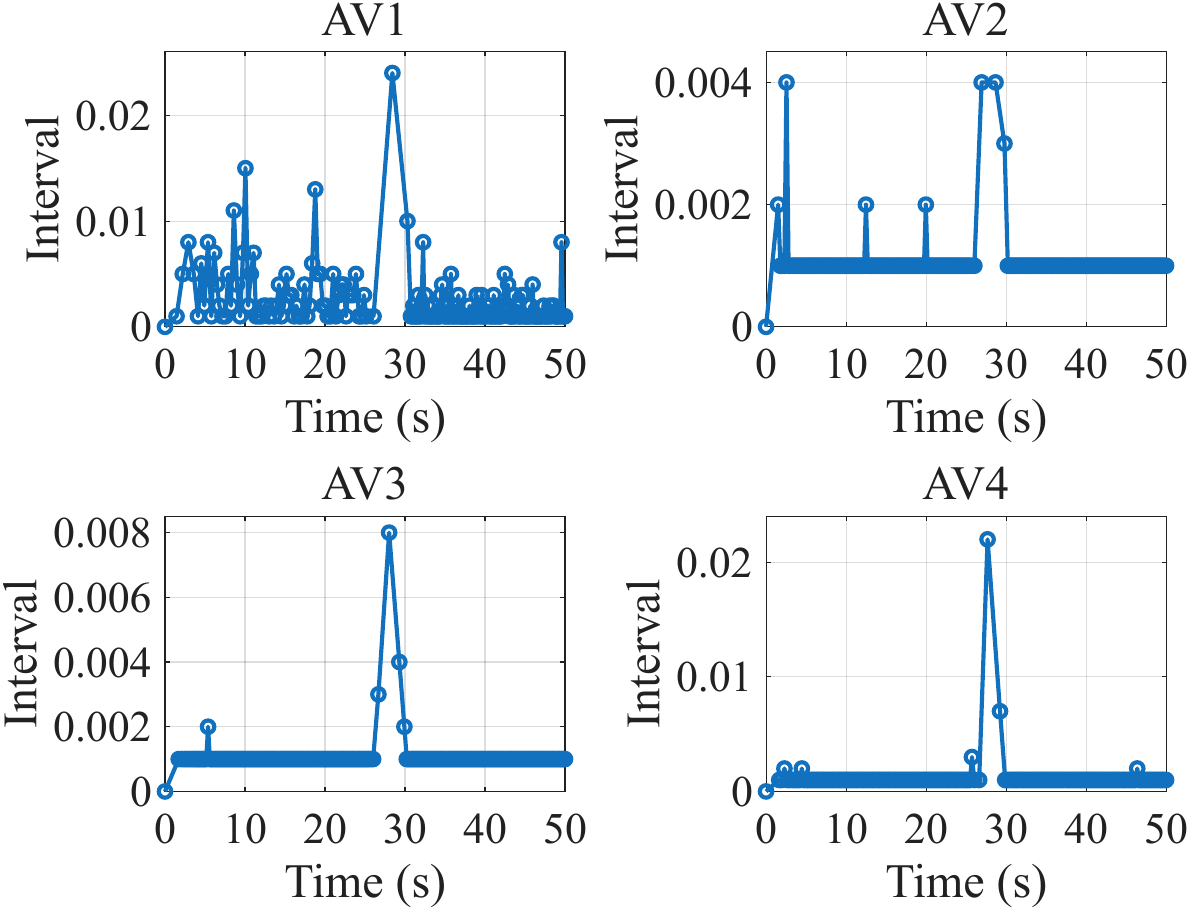}
    \caption{The Longitudinal update time interval of self-triggered control law
for the AVs in linear formation.}
    \label{Interval}
\end{figure}

The safe-distance results in Figure~\ref{tracking} (c) further confirm collision-free formation maintenance. Pairwise distances remain above safety thresholds with smooth temporal profiles, demonstrating robust spacing regulation. The evolution of self-triggered intervals in Figure~\ref{Interval} shows that the control-update frequency adapts to system dynamics: intervals shorten during transients and lengthen in steady states. Simulation statistics confirm that the proposed scheme lowers update events compared to continuous control while maintaining high tracking accuracy.

Overall, the results demonstrate that the proposed adaptive neural network controller with self-triggered updates achieves accurate trajectory tracking, stable formation maintenance, and reduced communication demands. The system exhibits fast transient response, negligible steady-state error, and reliable safety margins, validating its potential for AV platooning and cooperative formation control.

	\section{Conclusion}\label{sec:conclusion}
	This paper introduces an observer-based self-triggered adaptive control framework to tackle the consensus tracking problem of nonlinear MASs under communication constraints and uncertain dynamics. By integrating neural networks, a sampled-data observer, and a self-triggered control mechanism, the proposed approach alleviates the reliance on continuous communication and precise model constants. The theoretical analysis guaranties the uniform ultimate boundedness of all system signals while ensuring the asymptotic convergence of tracking errors to a neighborhood around the origin. The self-triggered mechanism maintains a strictly positive minimum inter-event time, preventing Zeno behavior and optimizing communication efficiency. Simulation results confirm the robustness and practicality of the approach under real-world conditions.

\begin{credits}
\subsubsection{\ackname} This study did not receive funding.

\subsubsection{\discintname}
 The authors have no competing interests to declare that are
relevant to the content of this article. 

\end{credits}
%
%
%

\begin{thebibliography}{8}
        \bibitem{lemma1}
		C. L. P. Chen, G. X. Wen, Y. J. Liu, and Z. Liu, "Observer-based adaptive backstepping consensus tracking control for high-Order nonlinear semi-strict-feedback multiagent systems,"\emph{IEEE Trans. Cybern.}, vol. 46, no. 7, pp. 1591--1601, 2016.

        
        \bibitem{wang2025fixed} 
        Z. Wang, Y. Gao, A. I. Rikos, N. Pang, and Y. Ji, "Fixed-relative-switched threshold strategies for consensus tracking control of nonlinear multiagent systems," in \textit{Proc. IEEE Int. Conf. Control Autom. (ICCA)}, pp. 899–905, 2025.

        \bibitem{cyber} X. Wang, Z. Yin, Y. Lei, T. Huang, and J. Kurths, "Secure consensus for switched multiagent systems under DoS attacks: Hybrid event-triggered and impulsive control approach," \textit{IEEE Trans. Cybern.}, vol. 55, no. 5, pp. 2400-2410, May 2025.

        \bibitem{case} 
        Z. Wang, S. Piao, Y. Ji, X. Wang, and F. Tsung, "An Efficient Dual-Observer Method for Leader-Following Consensus Control of Multiagent Systems," \textit{2025 IEEE 21st Int. Conf. Autom. Sci. Eng. (CASE)}, pp. 3468-3473, 2025.
        
		\bibitem{intro6}
		L. Wang, X. Wang, and Z. Wang, "Event-triggered optimal tracking control for strict-feedback nonlinear systems with non-affine nonlinear faults," \textit{Nonlinear Dyn.}, vol. 112, no. 17, pp. 15413-15426, 2024.

		\bibitem{tnnls}
		N. Pang, X. Wang and Z. Wang, "Observer-Based Event-Triggered Adaptive Control for Nonlinear Multiagent Systems With Unknown States and Disturbances," \textit{IEEE Trans. Neural Netw. Learn. Syst.}, vol. 34, no. 9, pp. 6663-6669, Sept. 2023.

        \bibitem{eve3.2}
		Z. M. Wang, "Hybrid Event-triggered Control of Nonlinear System with Full State Constraints and Disturbance," \textit{36th Chinese Control Decis. Conf.}, pp. 2122-2127, 2024. 

        \bibitem{ni2024sliding} Z. Ni, F. Li, Y. Wang, and H. Shen, "Sliding-mode control for 2-D hidden Markov jump Roesser systems with partial information and its application in metal rolling process," \textit{IEEE Trans. Autom. Sci. Eng.}, vol. 22, pp. 6851–6859, Sep. 2024.

        \bibitem{li2024passivity} F. Li, Z. Ni, L. Su, J. Xia, and H. Shen, "Passivity-based finite-region control of 2-D hidden Markov jump Roesser systems with partial statistical information," \textit{Nonlinear Anal. Hybrid Syst.}, vol. 51, p. 101433, 2024.

        \bibitem{eve1}
		K. J. Åström and B. Bernhardsson, "Comparison of periodic and event based sampling for first-order stochastic systems," \textit{IFAC Proceedings Volumes}, vol. 32, no. 2, pp. 5006-5011, Jul. 1999.

        \bibitem{chufa}
		L. T. Xing, C. Y. Wen, Z. T. Liu, H. Y. Su, and J. P. Cai, "Event-Triggered Adaptive Control for a Class of Uncertain Nonlinear Systems," \textit{IEEE Trans. Auto. Cont.}, vol. 62, no. 4, pp. 2071-2076, 2017.

        \bibitem{lemma2}
		Z. Wang, X. Wang and N. Pang, "Adaptive Fixed-Time Control for Full State-Constrained Nonlinear Systems: Switched-Self-Triggered Case" \textit{ IEEE Trans. Circuits Syst. II Express Briefs}, vol. 71, no. 2, pp. 752-756, 2024.

        \bibitem{eve4}
		P. Tabuada, "Event-Triggered Real-Time Scheduling of Stabilizing Control Tasks," \textit{IEEE Trans. Autom. Control.}, vol. 52, no. 9, pp. 1680-1685, 2007.

        \bibitem{eve2}
		D. Theodosis and D. V. Dimarogonas, "Event-Triggered Control of Nonlinear Systems With Updating Threshold," \textit{IEEE Control Syst. Lett.}, vol. 3, no. 3, pp. 655-660, 2019.

        \bibitem{cogcomp2025} L. Meng, X. Wang, and Z. Wang, "Event-Triggered Optimized Control for Nonlinear Multiagent Systems via Reinforcement Learning Strategy," \textit{Cogn. Comput.}, vol. 17, no. 5, pp. 1-9, 2025.
        
		\bibitem{eve3.1}
		P. Elena, P. Romain, A. Daniele, N. Dragan and H. W. Maurice, "Decentralized event-triggered estimation of nonlinear systems,"\emph{Automatica}, vol. 160, pp. 111414, 2024.     

        \bibitem{eve5}
		A. Anta and P. Tabuada, "To Sample or not to Sample: Self-Triggered Control for Nonlinear Systems," \textit{IEEE Trans. Autom. Control.}, vol. 55, no. 9, pp. 2030-2042, 2010.

		\bibitem{rela2}
		Z. M. Wang, H. Wang, X. Wang, N. Pang and Q. Shi, "Event-Triggered Adaptive Neural Control for Full State-Constrained Nonlinear Systems with Unknown Disturbances" \textit{Cogn. Comput.}, vol. 16, no. 2, pp. 717--726, 2023.

        \bibitem{tsmc2025} X. Wang, S. Zhang, H. Li, W. Zhang, H. Li, and T. Huang, "Neuroadaptive Containment Control for Nonlinear Multiagent Systems With Input Saturation: An Event-Triggered Communication Approach," \textit{IEEE Trans. Syst. Man Cybern. Syst.}, vol. 55, no. 5, pp. 3163-3173, May 2025.
        
        \bibitem{arxiv} 
        Z. Wang, Y. Zhang, C. Zhao, and H. Yu, "Adaptive Event-triggered Formation Control of Autonomous Vehicles," \textit{arXiv preprint arXiv:2506.06746}, 2025.

        \bibitem{trc} Y. Xue, C. Wang, C. Ding, B. Yu, and S. Cui, "Observer-based event-triggered adaptive platooning control for autonomous vehicles with motion uncertainties," \textit{Transp. Res. Part C Emerg. Technol.}, vol. 159, pp. 104462, 2024.
        
\end{thebibliography}
%

\end{document}